\documentclass[11pt]{article}

\usepackage{amsmath}
\usepackage{graphicx}
\usepackage{amsfonts}
\usepackage{amssymb}
\usepackage{epsfig}
\usepackage{color}
\usepackage{psfrag}
\usepackage{epstopdf}

\newcommand{\EQ}{\begin{equation}}
\newcommand{\EN}{\end{equation}}
\newcommand{\be}{\begin{equation}}
\newcommand{\ee}{\end{equation}}
\newcommand{\bea}{\begin{eqnarray}}
\newcommand{\eea}{\end{eqnarray}}

\newcommand{\rd}{{\rm d}}

\begin{document} \setcounter{page}{0}
\topmargin 0pt
\oddsidemargin 5mm
\renewcommand{\thefootnote}{\arabic{footnote}}
\newpage
\setcounter{page}{0}
\topmargin 0pt
\oddsidemargin 5mm
\renewcommand{\thefootnote}{\arabic{footnote}}
\newpage
\begin{titlepage}
\begin{flushright}
\end{flushright}
\vspace{0.5cm}
\begin{center}
{\large {\bf Phase separation in a wedge. Exact results}}\\
\vspace{1.8cm}
{\large Gesualdo Delfino and Alessio Squarcini}\\
\vspace{0.5cm}
{\em SISSA -- Via Bonomea 265, 34136 Trieste, Italy}\\
{\em INFN sezione di Trieste}\\
\end{center}
\vspace{1.2cm}

\renewcommand{\thefootnote}{\arabic{footnote}}
\setcounter{footnote}{0}

\begin{abstract}
\noindent
The exact theory of phase separation in a two-dimensional wedge is derived from the properties of the order parameter and boundary condition changing operators in field theory. For a shallow wedge we determine the passage probability for an interface with endpoints on the boundary. For generic opening angles we exhibit the fundamental origin of the filling transition condition and of the property known as wedge covariance.
\end{abstract}
\end{titlepage}

\newpage
Interfacial phenomena at boundaries are a subject of both experimental and theoretical relevance which has received continuous and extensive interest in the last decades \cite{deGennes}-\cite{BEIMR}. An aspect particularly important for applications is that the structure and geometry of the substrate can alter the adsorption characteristics of a fluid in an important way (see \cite{DPR} for a review). Adsorption measurements can then be used, for example, to characterize fractally rough surfaces \cite{PWCK}, or the connectivity of porous substrates \cite{GGRS}.
The basic case of a wedge-shaped substrate \cite{RP} acquired special interest since thermodynamic arguments \cite{Hauge} indicated a specific relation with the adsorption properties of a completely flat surface: the wedge wetting (or filling) transition occurs at the temperature for which the contact angle of a fluid drop on a flat substrate equals the tilt angle $\alpha$ of the wedge, a circumstance that allows to regulate the transition temperature adjusting $\alpha$. The connections between adsorption characteristics for different opening angles are known as properties of {\it wedge covariance} \cite{APW,RP2,PR} and are experimentally testable \cite{BCM}.

The importance to progress from a thermodynamic to a statistical mechanical description is obvious. In two dimensions the essential role of fluctuations was established by the exact lattice results for the Ising model on the half phane \cite{Abraham_wetting,Abraham}, which provided essential support for heuristic statistical descriptions of the wetting of a flat boundary \cite{Fisher}. For the wedge geometry the existence of the filling transition was established for the Ising model on a planar lattice forming a right-angle corner \cite{AM}, but otherwise theoretical investigation in two and three dimensions has been based on effective interfacial Hamiltonian models \cite{APW,RP2,PR,RDN,PRW} or density functional methods \cite{MP}. 

In this paper we derive general exact results for phase separation in a two-dimensional wedge. This is achieved exploiting low energy properties of bulk two-dimensional field theory \cite{DV,DS2} together with a characterization of the operators responsible for the departure of an interface from a point on a boundary. For a shallow wedge we determine the exact passage probability of an interface with endpoints on the boundary. The theory provides a fundamental meaning to the contact angle and, for generic $\alpha$, yields the filling transition condition. More generally, wedge covariance is shown to originate from the properties of the boundary condition changing operators in momentum space.

We begin the derivation considering a two-dimensional system at a first order transition point, close enough to a second order transition to allow a continuous description in terms of a Euclidean field theory on the plane with coordinates $(x,y)$. We label by an index $a=1,\ldots,n$ the different coexisting phases, denote by $\sigma(x,y)$ the order parameter operator, and by $\langle\sigma\rangle_a$ its expectation value in phase $a$. The Euclidean field theory corresponds to the continuation to imaginary time $t=iy$ of a relativistic quantum theory in one space dimension, for which phase coexistence amounts to the presence of degenerate vacua $|\Omega_a\rangle$ associated to the different phases of the system. The elementary excitations are kink states $|K_{ab}(\theta)\rangle$ interpolating between different vacua $|\Omega_a\rangle$ and $|\Omega_b\rangle$; the rapidity $\theta$ parameterizes energy and momentum of these particles as $p^0=m_{ab}\cosh\theta$ and $p^1=m_{ab}\sinh\theta$, respectively, the kink mass $m_{ab}$ being inversely proportional to the correlation length. The relativistic dispersion relation $(p^0)^2-(p^1)^2=m_{ab}^2$ is preserved by Lorentz boosts $M_\Lambda$ with matrix elements\footnote{The same matrix acts on space-time coordinates $x^0=t$ and $x^1=x$.} $M_\Lambda^{00}=M_\Lambda^{11}=\cosh\Lambda$ and $M_\Lambda^{01}=M_\Lambda^{10}=\sinh\Lambda$, which shift rapidities by $\Lambda$; as a consequence, relativistic invariant quantities depend on rapidities differences.
If $H$ and $P$ are the Hamiltonian and momentum operators of the quantum system, a local operator $\Phi$ satisfies $\Phi(x,y)=e^{ixP+yH}\Phi(0,0)e^{-ixP-yH}$.

\begin{figure}[t]
\centering
\includegraphics[width=4cm]{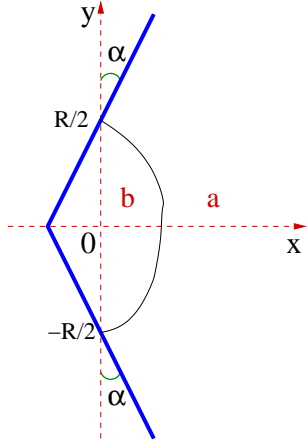}
\caption{Wedge geometry with boundary condition changing points at $(0,\pm R/2)$ and an interface running between them.}
\label{geometry}
\end{figure}

We want to consider the system in the wedge geometry of Fig.~\ref{geometry}, where the points $(0,\pm R/2)$ are boundary condition changing points. More precisely, a boundary field points in direction $a$ (resp. $b$) in order parameter space for $|y|>R/2$ (resp. $|y|<R/2$) on the wedge, inducing a symmetry breaking in the bulk. If we denote by a subscript $W_{aba}$ the statistical averages in the wedge geometry with these boundary conditions, the limit for $x\to\infty$ of the order parameter $\langle\sigma(x,y)\rangle_{W_{aba}}$ is expected to tend to $\langle\sigma\rangle_a$ if $R$ is finite, and to $\langle\sigma\rangle_b$ if $R$ is infinite. For $m_{ab}R$ large this should correspond to an interface running between the points $(0,\pm R/2)$, separating an inner phase $b$ from an outer phase $a$ (Fig.~\ref{geometry}), and whose average mid-point distance from the wedge diverges with $R$. 

To see how this emerges from the theory, consider first the case of a completely flat boundary (tilt angle $\alpha=0$). With a uniform boundary condition of type $a$ on the whole boundary we have a theory defined on the half plane $x\geq 0$ with a vacuum state that we denote $|\Omega_a\rangle_0$. The change of boundary condition at the point $(0,R/2)$ then corresponds to the insertion of an operator $\mu_{ab}(0,R/2)$ responsible for the emission from the boundary of kinks interpolating between phase $a$ and phase $b$. The simplest matrix element of such a boundary operator is
\EQ
{}_0\langle \Omega_a|\mu_{ab}(0,\frac{R}{2})|K_{ba}(\theta)\rangle_0=e^{-m\frac{R}{2}\cosh\theta}
{}_0\langle \Omega_a|\mu_{ab}(0,0)|K_{ba}(\theta)\rangle_0\equiv e^{-m\frac{R}{2}\cosh\theta}{\cal F}^\mu_0(\theta)\,,
\label{bff}
\EN
where we set $m_{ab}\equiv m$; $|K_{ba}(\theta)\rangle_0$ is an asymptotic kink state (outgoing if $\theta>0$) on the half plane and has energy $E_0+m\cosh\theta$, $E_0$ being the energy of the vacuum $|\Omega_a\rangle_0$. If we now move to a new frame through the Lorentz boost $M_\Lambda$, the kink rapidity becomes $\theta+\Lambda$, and the boundary moves in time. For a purely imaginary $\Lambda=-i\alpha$ the boost corresponds to a rotation in the Euclidean plane, and to a boundary forming an angle $\alpha$ with the $y$-axis. By relativistic invariance the emission amplitude ${\cal F}^\mu_\alpha(\theta-i\alpha)$ of the kink with rapidity $\theta-i\alpha$ from the moving boundary coincides with ${\cal F}^\mu_0(\theta)$, something that we can rewrite as
\EQ
{\cal F}^\mu_\alpha(\theta)={\cal F}^\mu_0(\theta+i\alpha)\,.
\label{Fcovariance}
\EN

The order parameter in the wedge for $|y|<R/2$ reads
\EQ
\langle \sigma(x,y) \rangle_{W_{aba}} = \frac{1}{\mathcal{Z}_{W_{aba}}}\,{}_\alpha\langle \Omega_{a} \vert \mu_{ab}(0,R/2)\sigma(x,y) \mu_{ba}(0,-R/2) \vert \Omega_{a} \rangle_{-\alpha}\,,
\label{op}
\EN
where $|\Omega_{a}\rangle_\alpha=M_{-i\alpha}|\Omega_{a}\rangle_0$ and
\EQ
\mathcal{Z}_{W_{aba}}={}_\alpha\langle \Omega_{a} \vert \mu_{ab}(0,R/2)\mu_{ba}(0,-R/2) \vert \Omega_{a} \rangle_{-\alpha}\,.
\label{pf}
\EN
We expand (\ref{pf}) inserting a complete set of asymptotic states between the two boundary condition changing operators, take the limit of large $mR$ in which the lightest (single kink) intermediate state dominates, and obtain
\EQ
\mathcal{Z}_{W_{aba}}\sim
\int_0^\infty\frac{\rd \theta}{2\pi}\,\mathcal{F}^\mu_\alpha(\theta)\mathcal{F}^\mu_{-\alpha}(\theta)\,\text{e}^{-mR(1+\frac{\theta^2}{2})}\,.
\label{partition01}
\EN
An emission amplitude ${\cal F}_0^\mu(0)\neq 0$ would mean that the kink has a probability of remaining stuck on the boundary, a possibility that we are excluding for the time being. Hence, on general analyticity grounds \cite{ELOP}, we have ${\cal F}_0^\mu(\theta)=c\,\theta+O(\theta^2)$, so that (\ref{Fcovariance}) gives
\EQ
{\cal F}_\alpha^\mu(\theta)\sim c(\theta+i\alpha)\,,\hspace{1cm}|\theta|,\,|\alpha|\ll 1\,,
\label{fmu}
\EN
and then
\EQ
\label{partition02}
\mathcal{Z}_{W_{aba}}\sim\frac{c^{2}\,\text{e}^{-mR}}{2\sqrt{2\pi}(mR)^{3/2}} \, (1 + mR \,\alpha^2)\,.
\EN
In a similar way, we evaluate (\ref{op}) expanding over intermediate states and taking the limits 
\EQ
mR\gg 1\,,\hspace{1cm}m\left(\frac{R}{2}-|y|\right)\gg 1\,,
\label{limits}
\EN
which project onto the single kink states. We then obtain
\EQ
\langle \sigma(x,y) \rangle_{W_{aba}}\sim \frac{e^{-mR}}{\mathcal{Z}_{W_{aba}}} \int_{-\infty}^{+\infty}\frac{\rd \theta_{1} \rd \theta_{2}}{(2\pi)^{2}}\mathcal{F}^\mu_{\alpha}(\theta_{1})\,\mathcal{M}_{\sigma}(\theta_{1}\vert\theta_{2})\,\mathcal{F}^\mu_{-\alpha}(\theta_{2})
\text{e}^{-\frac{m}{2}[(\frac{R}{2}-y)\theta_{1}^2+(\frac{R}{2}+y)\theta_{2}^2]+imx(\theta_{1}-\theta_{2})} \, ,
\label{op1}
\EN
\EQ
\mathcal{M}_{\sigma}(\theta_{1}\vert\theta_{2})\equiv\langle K_{ab}(\theta_{1})\vert \sigma(0,0) \vert K_{ba}(\theta_{2}) \rangle \,;
\label{ff}
\EN
evaluating the matrix element (\ref{ff}) on bulk kink states we imply that the leading boundary effects for large $mR$ are accounted for by the boundary condition changing operators inserted at the points $(0,\pm R/2)$. In the small rapidity limit which gives the leading contribution to the integral in (\ref{op1}) the matrix element of the order parameter operator takes the form \cite{DV,DS2}
\EQ
\mathcal{M}_{\sigma}(\theta_{1}\vert\theta_{2})\sim -\frac{i\Delta\langle\sigma\rangle}{\theta_{1}-\theta_{2}} + 2\pi\,\delta(\theta_{1}-\theta_{2}) \langle \sigma \rangle_{a}\,, \qquad \Delta\langle \sigma \rangle \equiv \langle \sigma \rangle_{a}-\langle \sigma \rangle_{b}\,.
\label{pole}
\EN
The delta function term is a disconnected part corresponding to the kink passing aside the operator without interacting; the pole comes from the fact that the operator sees different phases if the kink passes to its right or to its left. We use (\ref{fmu}) to evaluate (\ref{op1}) for small positive $\alpha$, but also differentiate with respect to $x$ in order to get rid of the pole in (\ref{pole}). The result is 
\bea 
\frac{\partial_{x}\langle \sigma(x,y)\rangle_{W_{aba}}}{\Delta\langle\sigma\rangle} &\sim & \frac{mc^{2}\text{e}^{-mR}}{\mathcal{Z}_{W_{aba}}} \int_{-\infty}^{+\infty}\frac{\rd \theta_{1} \rd \theta_{2}}{(2\pi)^{2}} \, (\theta_{1}+i\alpha)(\theta_{2}-i\alpha) \,\text{e}^{-\frac{m}{2}[(\frac{R}{2}-y)\theta_{1}^2+(\frac{R}{2}+y)\theta_{2}^2]+imx(\theta_{1}-\theta_{2})} \nonumber \\
&=&\frac{8\sqrt{2}}{\sqrt{\pi}\,\kappa^{3}} \left( \frac{m}{R} \right)^{3/2} \frac{\left(x+\frac{R\alpha}{2}\right)^{2}-(\alpha y)^{2}}{1+mR\alpha^{2}}\,\text{e}^{-\chi^{2}}  ,
\label{dop}
\eea
where we defined 
\EQ
\kappa = \sqrt{1-4y^{2}/R^{2}}\,,\hspace{1cm}\chi=\sqrt{\frac{2m}{R}}\,\frac{x}{\kappa}\,. 
\EN
Integrating back over $x$ with the asymptotic condition $\langle\sigma(+\infty,y)\rangle_{W_{aba}}=\langle\sigma\rangle_{a}$ gives
\EQ
\langle \sigma(x,y) \rangle_{W_{aba}}\sim\langle\sigma\rangle_{b}+\Delta\langle\sigma\rangle\Biggl[\text{erf}(\chi) -\frac{2}{\sqrt{\pi}}\,\frac{\chi + \sqrt{2mR}\,\frac{\alpha}{\kappa}}{1+mR\alpha^{2}}\,\text{e}^{-\chi^{2}} \Biggr]\, .
\label{op2}
\EN
For $\alpha=0$ and $\langle\sigma\rangle_{a}=-\langle\sigma\rangle_{b}$ this result coincides with that originally obtained in \cite{Abraham_wetting} from the exact lattice solution of the Ising model on the half plane.

\begin{figure}[t]
\centering
\includegraphics[width=7cm]{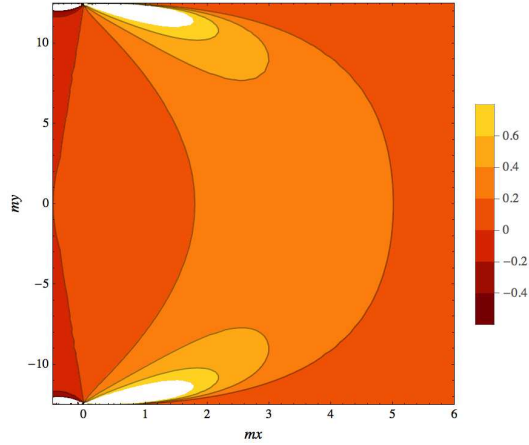}
\caption{Passage probability density $P(x,y)/m$ with $P(x,y)$ given by (\ref{dop}), $mR=25$ and $\alpha=0.04$. The leftmost contour line corresponds to $P(x,y)=0$, and then to the wedge.}
\label{wedge_pp}
\end{figure}

It was shown in \cite{DV,DS2,DS1} that the leading term in the large $mR$ (small rapidity) expansion corresponds to a sharp separation between pure phases, with subsequent terms accounting for the internal interface structure. In the present geometry the order parameter for sharp phase separation can be written as
\EQ
\langle \sigma(x,y) \rangle_{W_{aba}}\sim\langle \sigma \rangle_{a} \int_{\tilde{x}}^{x} \rd v \, P(v,y) + \langle \sigma \rangle_{b} \int_{x}^{\infty} \rd v \, P(v,y)\,,
\EN
where $P(v,y)$ is the probability that the interface passes in the interval $(v,v+dv)$ on the line of constant ordinate $y$, which intersects the wedge at $x=\tilde{x}(y)$. It follows that $P(x,y)$ coincides with (\ref{dop}). It can be checked that the consistency requirement $\int_{\tilde{x}}^{\infty}\rd x\,P(x,y)\sim 1$ implies $\sqrt{mR}\,\alpha\ll 1$. The result (\ref{dop}) shows in particular that $P(x,y)$ vanishes for $|y|=\frac{x}{\alpha}+\frac{R}{2}$, which in the small $\alpha$ approximation of the present computation are the coordinates of the wedge ($x\geq-R\alpha/2$). This shows that the properties (\ref{Fcovariance}), (\ref{fmu}) that we identified in momentum space for the matrix element of the boundary condition changing operator indeed lead to an impenetrable wedge in coordinate space. A plot of $P(x,y)$ is shown in Fig.~\ref{wedge_pp}. The average midpoint position of the interface reads
\EQ
\bar{x}=\int_{-R\alpha/2}^\infty\rd x\,x\,P(x,0)=\sqrt{\frac{2R}{\pi m}}\left[1+\sqrt{\frac{\pi mR}{2}}\,\alpha+O(mR\alpha^2)\right]\,.
\EN

The results (\ref{dop}), (\ref{op2}) apply to values of temperature (i.e. of  bulk correlation length $m^{-1}\propto (T_c-T)^{-\nu}$) for which the kink state is the lightest one entering the spectral decomposition of (\ref{op}). To discuss the situation in which this is not the case we start again from $\alpha=0$. For temperatures below a certain threshold $T_0<T_c$ the kink $K_{ba}(\theta)$ may form with the boundary a bound state $|\Omega'_a\rangle_0$ with energy $E'_0$, in which the phase $b$ forms a thin layer adsorbed on the boundary. As usual for stable bound states, this corresponds to a purely imaginary rapidity $\theta=i\theta_0$ of the kink, leading to a binding energy
\EQ
E'_0-E_0=m\cos\theta_0\,
\label{binding}
\EN
smaller than $m$. This boundary bound state is now the lightest contributing to (\ref{op}) and produces an order parameter equal to $\langle\sigma\rangle_a$ for $mx\gg 1$, no matter how large is $R$ \cite{DS1}. Since $m$ is the interfacial tension between the phases $a$ and $b$ \cite{DV}, and $E_0$ (resp. $E'_0$) corresponds to the tension between the boundary and phase $b$ (resp. $a$), (\ref{binding}) identifies $\theta_0$ as the contact angle of the phenomenological wetting theory. The usual relation $\theta_0(T_0)=0$ characterizing the wetting transition temperature $T_0$ corresponds to the unbinding threshold for the kink.

\begin{figure}[t]
\centering
\includegraphics[width=6cm]{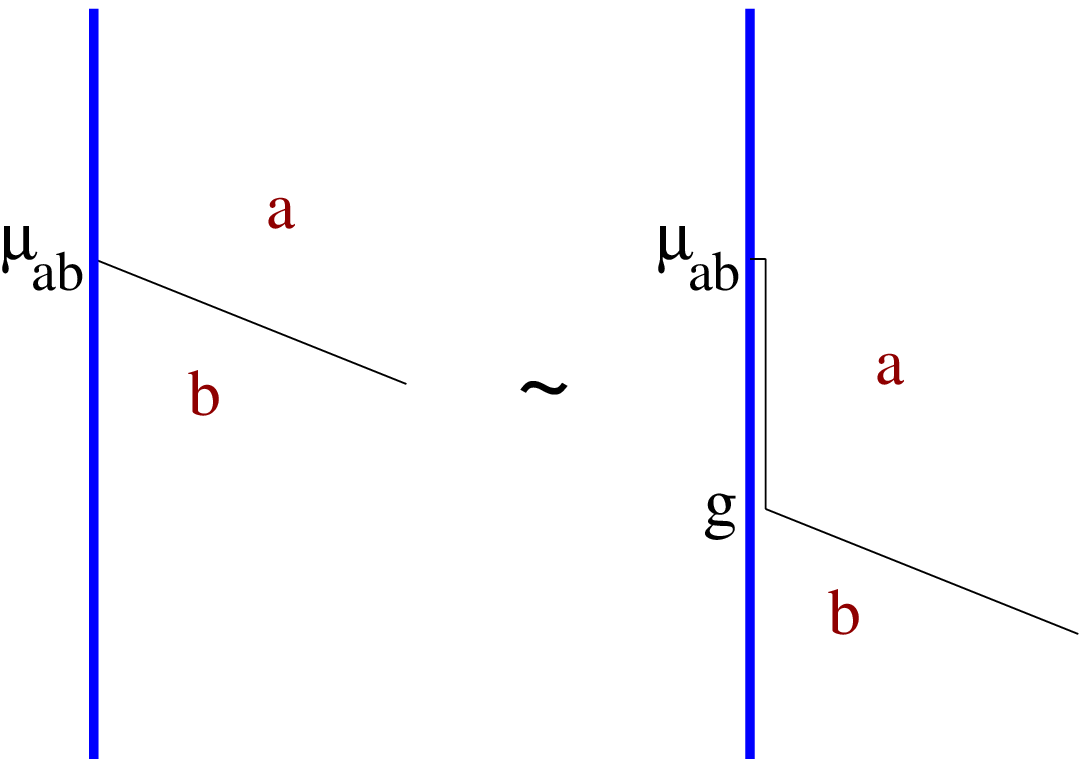}
\caption{Illustration of equation (\ref{ff_pole}).}
\label{wedge_pole}
\end{figure}

Bound states manifest in matrix elements as poles in the physical region of kinematical variables \cite{ELOP}. For the matrix element (\ref{bff}) this physical region corresponds to the strip Im\,$\theta\in(0,\pi)$, and the pole induced by the boundary bound state takes the form (see Fig.~\ref{wedge_pole})
\EQ
{\cal F}_0^\mu(\theta)={}_0\langle\Omega_a|\mu_{ab}(0,0)|K_{ba}(\theta)\rangle_0\sim\frac{ig}{\theta-i\theta_0}\,{}_0\langle\Omega_a|\mu_{ab}(0,0)|\Omega'_a\rangle_0\,,\hspace{1cm}\theta\sim i\theta_0\,,
\label{ff_pole}
\EN
with $g$ a coupling measuring the strength of the bound state. It then follows from (\ref{Fcovariance}) that the pole of ${\cal F}_\alpha^\mu(\theta)$ is located at $\theta=i(\theta_0-\alpha)$. Since the kink energy is always $m\cosh\theta$ and the unbinding threshold remains at $\theta=0$, the filling transition temperature $T_\alpha$ is determined by the condition
\EQ
\theta_0(T_\alpha)=\alpha\,.
\label{filling}
\EN
For $\theta_0<\alpha$ the pole is located at Im\,$\theta<0$, namely outside the physical strip allowed for bound states; in such a case the kink is unbounded and phase $b$ fills the wedge. The condition (\ref{filling}) is that obtained in the phenomenological framework \cite{Hauge}, and follows here from the exact statistical theory. Notice that while (\ref{dop}) and (\ref{op2}) rely on (\ref{fmu}), and then on small $\alpha$, (\ref{Fcovariance}) and (\ref{filling}) are general. Equation (\ref{Fcovariance}), in particular, encodes the essence of what is called wedge covariance.

For the scaling Ising model on the half plane with a boundary magnetic field $h$, the scattering amplitude off the boundary is known exactly \cite{GZ}, and exhibits a boundary bound state pole corresponding to $1-\sin\theta_0=\frac{h^2}{2m}=\frac{T_c-T_0(h)}{T_c-T}$; the last equality follows from $\nu_{Ising}=1$ and holds in the scaling limit.

In summary, we constructed the exact theory of phase separation in a two-dimensional wedge and derived from it the filling transition condition and the origin of wedge covariance. The passage probability for an interface with endpoints on the wedge has also been determined for small tilt angles in the unbound regime.


\begin{thebibliography}{99}
\bibitem{deGennes} P.G. De Gennes, Rev. Mod. Phys. 57 (1985) 827.
\bibitem{Dietrich} S. Dietrich, in Phase Transitions and Critical Phenomena, edited by C. Domb and J.L. Lebowitz,  Vol. 12, p.~1, Academic Press, London, 1988.
\bibitem{Schick} M. Schick, in Liquids at Interfaces, edited by J. Chavrolin, J.-F. Joanny and J. Zinn-Justin, p. 415, Elsevier, Amsterdam, 1990.
\bibitem{BEIMR} D. Bonn, J. Eggers, J. Indekeu, J. Meunier and E. Rolley, Rev. Mod. Phys. 81 (2009) 739.
\bibitem{DPR} S. Dietrich, M.N. Popescu and M. Rauscher, J. Phys.: Condensed Matter 17 (2005) S577.
\bibitem{PWCK} P. Pfeifer, Y.J. Wu, M.W. Cole and J. Krim, Phys. Rev. Lett. 62 (1989) 1997.
\bibitem{GGRS} L.D. Gelb, K.E. Gubbins, R. Radhakrishnan and M. Sliwinska-Bartkowiak, Rep. Prog. Phys. 62 (1999) 1573.
\bibitem{RP} C. Rasc\'on and A.O. Parry, Nature (London) 407 (2000) 986.
\bibitem{Hauge} E.H. Hauge, Phys. Rev. A 46 (1992) 4994. 
\bibitem{APW} D.B. Abraham, A.O. Parry and A.J. Wood, Europhys. Lett. 60 (2002) 106.
\bibitem{RP2} C. Rasc\'on and A.O. Parry, Phys. Rev. Lett. 94 (2005) 096103.
\bibitem{PR}  A.O. Parry and C. Rasc\'on, J. of Chem. Phys. 132 (2010) 204704.
\bibitem{BCM} L. Bruschi, A. Carlin and G. Mistura, Phys. Rev. Lett. 89 (2002) 166101.
\bibitem{Abraham_wetting} D.B. Abraham, Phys. Rev. Lett. 44 (1980) 1165.
\bibitem{Abraham} D.B. Abraham, in Phase Transitions and Critical Phenomena, edited by C. Domb and J.L. Lebowitz,  Vol. 10, p.~1, Academic Press, London, 1986.
\bibitem{Fisher} M.E. Fisher, J. Stat. Phys. 34 (1984) 667.
\bibitem{AM} D.B. Abraham and A. Maciolek, Phys. Rev. Lett. 89 (2002) 286101.
\bibitem{RDN} K. Rejmer, S. Dietrich and M. Napiorkowski, Phys. Rev. E 60 (1999) 4027.
\bibitem{PRW} A.O. Parry, C. Rasc\'on and and A.J. Wood, Phys. Rev. Lett. 83 (1999)  5535.

A.O. Parry, C. Rasc\'on and and A.J. Wood, Phys. Rev. Lett. 85 (2000) 345.
\bibitem{MP} A. Malijevsk\'y and A.O. Parry, Phys. Rev. Lett. 110 (2013) 166101.\bibitem{DV} G.~Delfino and J.~Viti, {J. Stat. Mech.} (2012) P10009.
\bibitem{DS2} G. Delfino and A. Squarcini, Annals of Physics 342 (2014) 171. 
\bibitem{ELOP} R.J. Eden, P.V. Landshoff, D.I. Olive, J.C. Polkinghorne, The Analitic S-Matrix, Cambridge University Press, 1966.
\bibitem{DS1} G.~Delfino and A. Squarcini, J. Stat. Mech. (2013) P05010.
\bibitem{GZ} S. Ghoshal and A.B. Zamolodchikov, Int. J. Mod. Phys. A9 (1994) 3841; Erratum, ibidem A9 (1994) 4353.



\end{thebibliography}
\end{document}